\documentclass[12pt,aps,prd,preprint,tightenlines,superscriptaddress,showpacs,nofootinbib]
{revtex4-1}
\usepackage{epsfig}
\usepackage{amssymb,amsmath}
\usepackage{color}
\usepackage[utf8]{inputenc} 

\newcommand{\bea}{\begin{eqnarray}}
\newcommand{\eea}{\end{eqnarray}}

 \makeatletter
\def\alt{\mathrel{\mathpalette\gl@align<}}
\def\agt{\mathrel{\mathpalette\gl@align>}}
\def\gl@align#1#2{\lower.6ex\vbox{\baselineskip\z@skip\lineskip\z@
\ialign{$\m@th#1\hfil##\hfil$\crcr#2\crcr\sim\crcr}}} \makeatother

\begin{document}
%
\vspace*{1.0cm}

\begin{center}
\baselineskip 20pt 
{\Large\bf 
Braneworld Cosmological Effect on Freeze-in Dark Matter Density 
and Lifetime Frontier
}
\vspace{1cm}

{\large 
Victor Baules, Nobuchika Okada and  Satomi Okada
}
\vspace{.5cm}

{\baselineskip 20pt \it
Department of Physics and Astronomy, University of Alabama, Tuscaloosa, AL35487, USA
} 

\vspace{.5cm}

\vspace{1.5cm} {\bf Abstract}
\end{center}

In the 5-dimensional braneworld cosmology, the Friedmann equation 
  of our 4-dimensional universe on a brane is modified at high temperatures 
  while the standard Big Bang cosmology is reproduced at low temperatures. 
Based on two well-known scenarios,  
  the Randall-Sundrum and Gauss-Bonnet braneworld cosmologies, 
  we investigate the braneworld cosmological effect on the relic density 
  of a non-thermal dark matter particle 
  whose interactions with the Standard Model particles are so weak that 
  its relic density is determined by the freeze-in mechanism. 
For dark matter production processes in the early universe,    
  we assume a simple scenario with a light vector-boson mediator for the dark matter particle
  to communicate with the Standard Model particles. 
We find that the braneworld cosmological effect can dramatically alters  
  the resultant dark matter relic density from the one in the standard Big Bang cosmology.  
As an application, we consider a right-handed neutrino dark matter 
  in the minimal $B-L$ extended Standard Model with a light $B-L$ gauge boson ($Z^\prime$) as a mediator. 
We find an impact of the braneworld cosmological effect 
  on the search for the long-lived $Z^\prime$ boson at the planned/proposed Lifetime Frontier experiments.  

\thispagestyle{empty}

\newpage

\addtocounter{page}{-1}

\baselineskip 18pt

\section{Introduction} 
Based on various cosmological and astronomical observations, 
  including very precise measurements of the cosmic microwave background anisotropy, 
  the so-called $\Lambda$CDM cosmological model has been established and 
  the abundance of the (cold) dark matter (DM) is estimated as \cite{Aghanim:2018eyx}
\begin{equation}
\Omega_{\rm DM} h^2 \simeq 0.12.
\end{equation}
Since the Standard Model (SM) of particle physics has no suitable candidate 
  for the cold DM particle, new physics beyond the SM is required to supplement 
  the SM with a DM candidate. 
Many models of new physics have been proposed to incorporate various DM candidates. 
For a review, see Ref.~\cite{Bertone:2004pz}.

Among several possibilities, the most popular DM scenario is the thermal DM, 
  in which a DM particle was in thermal equilibrium in the early universe
  and its relic density at present is determined by the freeze-out mechanism \cite{Lee:1977ua}. 
In this scenario, we evaluate the DM relic density 
  by solving the Boltzmann equation \cite{Kolb:1990vq}, 
\bea 
\frac{d Y}{d x}
= -\frac{s(T=m)}{H(T=m)} \,  \frac{\langle\sigma v_{\rm rel} \rangle}{x^2} \, (Y^2-Y_{EQ}^2) ,
\label{Y:Boltzmann1}
\eea 
where $x=m/T$ is the ratio between the DM mass ($m$) and the temperature of the universe ($T$), 
  $Y=n/s$ is the yield defined by the ratio of the DM number density $(n(T))$ to the entropy density 
  of the universe $s(T) = (2 \pi^2/45) g_* T^3$ with $g_*$ being the effective total number of relativistic degrees of freedom, 
  $H(T)=\sqrt{\pi^2 g_*/90} \, (T^2/M_P)$ is the Hubble parameter 
  with the reduced Planck mass of $M_P=2.43 \times 10^{18}$ GeV, 
  $\langle\sigma v_{\rm rel}\rangle$ is the thermal-averaged DM pair annihilation cross section ($\sigma$) times 
  DM relative velocity ($v_{\rm rel}$), 
  and $Y_{EQ}$ is the yield for the DM in thermal equilibrium. 
Once a DM model is fixed, we can calculate $\langle\sigma v_{\rm rel} \rangle$ as a function of $x$
  and evaluate the yield of the thermal DM particle at present $Y(x \to \infty)$ 
  by solving the Boltzmann equation with the initial condition of $Y=Y_{EQ}$ for $x \ll 1$. 
In the freeze-out mechanism, we can approximate $Y(\infty)$ by $Y(\infty) \simeq Y(x_f)$ 
  at the freeze-out temperature $x_f=m/T_f$.   
The relic density of the DM particle is given by 
\bea
 \Omega_{\rm DM} h^2 = \frac{m \, Y(\infty) \, s_0}{\rho_c/h^2}, 
\eea
where $s_0=2890/{\rm cm}^3$ is the entropy density of the present universe, and 
  $\rho_c/h^2=1.05 \times 10^{-5}$ GeV/cm$^3$ is the critical density.

It is well-known that the observed DM relic density of $\Omega_{\rm DM} h^2 \sim 0.1$ 
  is obtained by $\langle\sigma v_{\rm rel}\rangle \sim 1$ pb, almost independently of the DM mass. 
This cross section at the electroweak scale implies the possibility of directly detecting a DM particle 
  through its elastic scattering off of nucleons. 
Despite many experimental efforts, evidence of the DM particle has not been observed yet. 
For example, the most stringent upper bound 
  of around $4 \times 10^{-11}$ pb for a DM particle with a 30 GeV mass
  has been set by the XENON 1T experiment \cite{Aprile:2018dbl}. 
Although the discovery of a DM particle may be around the corner, 
  the null result motivates us to consider the possibility that the interaction of a DM particle
  with SM particles is extremely weak. 
If this is the case, a DM particle has never been in thermal equilibrium with the plasma of the SM particles
  in the history of the universe. 
In such a case, the DM relic density is determined by the freeze-in mechanism \cite{McDonald:2001vt, Hall:2009bx} 
  (see Ref.~\cite{Bernal:2017kxu} for a review), 
  assuming a vanishing initial DM density at the reheating after inflation. 
The freeze-in DM scenario has attracted a lot of attention recently.

We also employ the Boltzmann equation to evaluate the DM relic density for the freeze-in DM particle. 
The difference from the thermal DM scenario is only the boundary condition, 
   namely, $Y(x_{RH})=0$ instead of $Y(x \ll 1)=Y_{EQ}$, 
   where $x_{RH}  \equiv m/T_{RH} \ll 1$ with the reheating temperature $T_{RH}$ after inflation. 
Since $Y$ never reaches $Y_{EQ}$ because of the extremely weak DM interactions with the SM particles, 
   the Boltzmann equation has the approximate form: 
\bea 
\frac{d Y}{d x} \simeq \frac{s(m)}{H(m)} \,  \frac{\langle\sigma v_{\rm rel} \rangle}{x^2} \, Y_{EQ}^2
   \simeq  0.698 \,  \frac{g_{DM}^2}{g_*^{3/2}} \, m \, M_P \,  \frac{\langle\sigma v_{\rm rel} \rangle}{x^2}.  
\label{Y:Boltzmann2}
\eea 
In the last expression, we have used $n_{EQ}= (g_{DM}/\pi^2) T^3$ (for $T > m$) with $g_{DM}$ 
  being the DM internal degrees of freedom.  
Here, note that $\langle\sigma v_{\rm rel} \rangle Y_{EQ}^2$ corresponds to the creation rate 
  of a pair of DM particles by the thermal plasma because it balances with the DM annihilation rate
  when the DM particle is in thermal equilibrium.  
For a given $\langle\sigma v_{\rm rel} \rangle$ as a function of $x$, it is easy to solve the Boltzmann equation, 
  and we estimate $Y(\infty) \simeq Y(x=1)$. 
Note that we integrate the Boltzmann equation until $x=1$ 
  since the production of DM particles from thermal plasma
  stops around $x\sim1$, or equivalently, $T \sim m$ by kinematics.   
As a simple DM scenario, we consider the case that the DM particle communicates with the SM particles    
  through a light vector-boson mediator. 
In this case, we may express the DM creation/annihilaton cross section as 
\bea
   \langle\sigma v_{\rm rel} \rangle = \frac{g_V^4}{128 \pi} \, \frac{x^2}{m^2}, 
 \label{Xsec}  
\eea
  where $g_V$ is a coupling of the vector-boson. 
Using this concrete form, one can easily solve Eq.~(\ref{Y:Boltzmann2}) to arrived at 
\bea
  \Omega_{\rm DM} h^2 =\frac{m \, Y(x=\infty) \, s_0}{\rho_c/h^2} \simeq \frac{m \, Y(x=1) \, s_0}{\rho_c/h^2} 
  \simeq 1.16 \times 10^{24}  \, \frac{g_{DM}^2}{g_*^{3/2}}  \, g_V^4. 
\label{Omega}
\eea
Interestingly, the resultant relic density is independent of the DM mass. 
For example, by using $g_{DM}=2$ and $g_*=106.75$ for the SM particle plasma, 
  we find $g_V = 2.31 \times 10^{-6}$  to reproduce $\Omega h^2 = 0.12$.

Since the discovery of the ``brane'' in string theories \cite{Polchinski:1996na}
  the braneworld scenarios have attracted lots of attention as phenomenological models, 
  in which the SM particles are confined on a ``3-brane'' while gravity resides in the bulk space.  
The braneworld cosmology based on the model first proposed by Randall and Sundrum (RS) \cite{Randall:1999vf} 
  (the so-called RS II model) has been intensively investigated (see Ref.~\cite{Langlois:2002bb} for a review). 
It has been found \cite{Binetruy:1999ut,Binetruy:1999hy, Shiromizu:1999wj, Ida:1999ui} 
  that the Friedmann equation in the RS cosmology 
  leads to a non-standard expansion law at high temperatures
  while the standard Big Bang cosmology is reproduced at low temperatures.
The RS II model can be extended by adding the Gauss-Bonnet (GB) invariant 
  \cite{Kim:1999dq,  Kim:2000pz, Nojiri:2002hz, Lidsey:2002zw}, 
  and the Friedmann equation for the GB braneworld cosmology has been found \cite{Charmousis:2002rc,  Maeda:2003vq},   
  which is quite different from the one in the RS braneworld cosmology.  
   
Since the Hubble parameter is involved in the Boltzmann equation, 
  the DM relic density depends on the expansion law of the early universe. 
Hence, the non-standard evolution of the universe can significantly alter 
  the resultant DM density from that found in the standard Big Bang cosmology.

The RS braneworld cosmological effect on the thermal DM physics has been investigated in detail 
  \cite{Okada:2004nc, Nihei:2004xv, Nihei:2005qx, AbouElDahab:2006glf,
     Panotopoulos:2007fg, Okada:2007na, Kang:2008jq, Guo:2009nt, Bailly:2010hh, 
      Meehan:2014zsa, Iminniyaz:2018das}, and  it has been shown 
      that the resultant DM density can be considerably enhanced. 
On the other hand, the GB braneworld cosmological effect has been shown
  to considerably reduce the thermal DM density \cite{Okada:2009xe, Meehan:2014bya}. 
   
In this paper, we investigate the braneworld cosmological effect on the non-thermal DM scenario 
  in which the DM relic density is determined
  by the freeze-in mechanism (for a related work, see Ref.~\cite{Bernal:2019mhf}). 
In particular, we focus on a freeze-in DM which communicates with the SM particles 
  through a light vector-boson mediator. 
For the two typical scenarios, namely, the RS and the GB braneworld cosmologies, 
  we evaluate the DM relic density under the non-standard evolution of the universe. 
We will find interesting RS and GB braneworld cosmological effects. 
As an application of our findings, we consider a right-haded neutrino (RHN) DM scenario 
  \cite{Okada:2010wd} (see also Ref.~\cite{Anisimov:2008gg})
  in the context of the minimal $B-L$ extended SM \cite{Davidson:1978pm, Mohapatra:1980qe, Marshak:1979fm,
          Wetterich:1981bx, Masiero:1982fi,  Mohapatra:1982xz, Buchmuller:1991ce}, 
  where the RHN DM communicates with the SM particles through a light $B-L$ gauge boson ($Z^\prime$). 
Because of a small $B-L$ gauge coupling for the freeze-in mechanism, 
   the $Z^\prime$ boson can be long-lived. 
When the $Z^\prime$ boson mass lies in the range of 10 MeV $\lesssim m_{Z^\prime} \lesssim 1$ GeV, 
   the planned/proposed Lifetime Frontier experiments can explore such a long-lived $Z^\prime$ boson. 
We find an impact of the braneworld cosmological effect on such experiments.   
  
This paper is organized as follows: 
In the next section, we give a brief review on the RS and GB braneworld cosmologies. 
In Sec.~\ref{sec:3}, we consider a freeze-in DM scenario with a light vector-boson mediator
   in the braneworld cosmology and discuss how the braneworld effect alters the resultant DM density 
   from the one in the standard Big Bang cosmology. 
In Sec.~\ref{sec:4}, we apply our findings in Sec.~\ref{sec:3} 
   to the so-called $Z^\prime$-portal RHN DM scenario in the context of the minimal $B-L$ model, 
   and identify the parameter region to reproduce the observed DM density. 
We also discuss the search for a long-lived $B-L$ gauge boson     
   by the planned/proposed experiments at Lifetime Frontier 
   and point out an impact of the braneworld cosmological effect on the search. 
Sec.~\ref{sec:5} is devoted to conclusions.

\section{Braneworld cosmologies}
\label{sec:2}
In this section, we give a brief review on two typical braneworld cosmologies: the RS cosmology 
  and the GB cosmology. 
In both scenarios, the Standard Big Bang cosmology is reproduced at low temperatures 
  while the evolution of the universe obeys non-standard law at high temperatures. 
Considering this fact, we may parametrize a modified Friedmann equation 
  in a braneworld cosmology as 
\bea
 H = H_{\rm st}(T)  \times F(T), 
\label{eq:parametrize1}
\eea
where $H_{\rm st}=\sqrt{\pi^2 g_*/90} \, (T^2/M_P)$ is the Hubble parameter in the standard Big Bang cosmology.  
Since the standard Big Bang cosmology must be reproduced at low temperatures, 
   we express $F(T)$ as $F(T/T_t)$ with ``transition temperature" ($T_t$) 
   at which the modified expansion law approaches the standard expansion law, 
   namely,  $F(T/T_t) =F(x_t/x) \to 1$ for $T <T_t$, where $x_t=m/T_t$.  
For concreteness, let us assume the following form: 
\bea
  F (T/T_t) =   \left( \frac{T}{T_t} \right)^\gamma =\left(  \frac{x_t}{x} \right)^\gamma  
\label{eq:parametrize2}
\eea
  for $T/T_t > 1$ with a real parameter $\gamma $.  
This parameterization turns out to be a very good approximation for both of the RS and GB braneworld cosmologies. 
As we will see, $\gamma= 2$ corresponds to the RS braneworld cosmology, 
   while $\gamma= -2/3$ to the GB braneworld cosmology.

\subsection{RS cosmology}
In the RS cosmology, the Friedmann equation for a spatially flat universe 
  is found to be \cite{Binetruy:1999ut,Binetruy:1999hy, Shiromizu:1999wj, Ida:1999ui} 
\begin{equation}
H^2 = \frac{\rho}{3 M_P^2} \left(1+\frac{\rho}{\rho_{\rm RS}} \right) ,
\label{Eq.RSH}
\end{equation}
where $\rho$ is the energy density of the universe, and
\begin{eqnarray}
 \rho_{\rm RS} = 12 \,  \frac{M_5^6}{M_P^2},
\label{rho_0}
\end{eqnarray}
   with $M_5$ being the 5-dimensional (5D) Planck mass. 
Here, we have set the model parameters to make the 4D cosmological constant 
   and the so-called dark radiation \cite{Ichiki:2002eh} vanishing.    
Note that the Friedmann equation of the standard Big Bang cosmology is reproduced 
    at low energies (temperatures) such that $\rho/\rho_{\rm RS} \ll 1$.  
A lower bound on $\rho_{\rm RS}^{1/4} \gtrsim 1.3$ TeV, or equivalently, $M_5 \gtrsim 1.1 \times 10^8$ GeV
   was obtained in Ref.~\cite{Randall:1999vf} from the precision measurements of the gravitational law in sub-millimeter range.

Let us consider the radiation dominated era in the early universe, 
   where the temperature is so high that $\rho/\rho_{\rm RS} \gg 1$. 
Then, the Friedmann equation of  Eq.~(\ref{Eq.RSH}) can be approximated by
\bea
  H   \simeq H_{\rm st} \, \sqrt{\frac{\rho}{\rho_{\rm RS}}} = H_{\rm st} \times \left( \frac{x_t}{x} \right)^2
\eea 
   where we have used $\rho/\rho_{\rm RS} = (T/T_t)^4 = (x_t/x)^4$. 
Hence, we find $\gamma=2$ in Eq.~(\ref{eq:parametrize2}) for the RS cosmology ($T \gg T_t$).

\subsection{GB cosmology}
The RS II model can  be generalized by adding higher curvature terms
  \cite{Kim:1999dq,  Kim:2000pz, Nojiri:2002hz, Lidsey:2002zw}. 
Among various possibilities, the GB invariant is of particular interests in 5D 
   since it is a unique nonlinear term in curvature yielding the gravitational field equations at the second order. 
The action of the RS II model is extended by adding  the GB invariant  \cite{Kim:1999dq,  Kim:2000pz, Nojiri:2002hz, Lidsey:2002zw}: 
\bea
 {\cal S} &=& \frac{1}{2\kappa_5^2} \int d^5x 
 \sqrt{-g_5}
 \left[
 - 2 \Lambda_5+ {\cal R} + 
 \alpha \left( {\cal R}^2 -4 {\cal R}_{ab} {\cal R}^{ab} 
 + {\cal R}_{a b c d}{\cal R}^{a b c d} \right) \right] \nonumber \\
&-& \int_{brane} d^4x 
 \sqrt{-g_4} \left( m_\sigma^4 + {\cal L}_{matter} \right), 
\label{GBaction}
\eea
where indices $a,b,c,d$ run $0$ to $4$, $\kappa_5^2=8\pi/M_5^3$, $ m_\sigma^4 >0 $ is a brane tension, 
  $ \Lambda_5 <0 $ is the bulk cosmological constant, 
  and a $Z_2$-parity across the brane in the bulk is imposed. 
The limit $\alpha \to 0$ recovers the RS II model. 

The Friedmann equation on the spatially flat brane has been found to be \cite{Charmousis:2002rc,  Maeda:2003vq}
\bea 
 \kappa_5^2(\rho + m_\sigma^4) = 
  2 \mu \sqrt{1+\frac{H^2}{\mu^2}}
  \left( 3 - \beta +2 \beta \frac{H^2}{\mu^2} \right) ,  
\label{GBFriedmann1} 
\eea
where $\beta = 4 \alpha \mu^2 = 1-\sqrt{1 + 4 \alpha \Lambda_5/3}$. 
The model involves four free parameters, $\kappa_5$, $m_\sigma$, $\mu$ and $\beta$. 
We impose two phenomenological requirements: 
   (i) the Friedmann equation of the standard Big Bang cosmology must be reproduced 
   at low energies $H^2/\mu^2 \ll 1$; 
   (ii) 4D cosmolgical constant is approximately zero.  
These requirements lead to the following two conditions: 
\bea 
 \kappa_5^2 m_\sigma^4 = 2 \mu (3-\beta), \; \; \;  \frac{1}{M_P^2} 
  = \frac{\mu}{1+\beta} \kappa_5^2 .
\label{relations} 
\eea
In general, the GB cosmology has three epochs in its evolution \cite{Lidsey:2003sj}: 
The universe obeys the standard expansion law at low energies (standard epoch). 
At middle energies (RS epoch), the RS cosmology is approximately realized. 
At high energies (GB epoch), the Friedmann equation is approximately expressed as
\bea 
 H \simeq 
 \left(  \frac{1+\beta}{4 \beta}  \, \frac{\mu}{M_P^2}  \, \rho  \right)^{1/3}.  
\eea
For a special value of $\beta = 0.151$ which satisfies the equation, 
  $ 3 \beta^3 -12 \beta^2 +15 \beta -2 =0$, 
   the RS epoch collapses \cite{Lidsey:2003sj}. 
Since we are interested in the GB epoch, we fix $\beta = 0.151$. 
Hence, we can parametrize the Friedman equation in the GB epoch as 
\bea
   H \simeq  \left(  \frac{1+\beta}{4 \beta}  \, \frac{\mu}{M_P^2}  \, \rho  \right)^{1/3} 
    = H_{\rm st}  \times  \left( \frac{\rho}{\rho_{\rm GB}}  \right)^{-1/6}
    = H_{\rm st}  \times  \left( \frac{x_t}{x}  \right)^{-2/3}, 
\eea
   where $\rho_{\rm GB}=3^6 \left(\frac{1+\beta}{4 \beta} \mu M_P \right)^2$. 
Thus, taking $\gamma=-2/3$ in Eq.~(\ref{eq:parametrize2}) corresponds to the GB cosmology for $T \gg T_t$.

\section{Freeze-in dark matter in Braneworld cosmologies}
\label{sec:3}
Now we are ready to investigate the braneworld cosmological effect on the freeze-in DM scenario. 
Our crucial assumption is that the DM mass is larger than the transition temperature ($m > T_t$), 
   so that DM production from the SM thermal plasma ends 
   before the evolution of the universe transitions to the standard Big Bang cosmology. 
To evaluate the freeze-in DM density, we solve the Boltzmann equation. 
All the braneworld cosmological effect is encoded in the modification of the Hubble parameter 
   given by Eq.~(\ref{eq:parametrize1}) with Eq.~(\ref{eq:parametrize2}). 
Therefore, we obtain a modified Boltzmann equation of the form:  
\bea 
\frac{d Y}{d x}= \frac{s(m)}{H_{\rm st}(m)} \,  \frac{\langle\sigma v_{\rm rel} \rangle}{F(x_t/x) \, x^2} \, Y_{EQ}^2
   \simeq  0.698 \,  \frac{g_{DM}^2}{g_*^{3/2}} \, m \, M_P \,  \frac{\langle\sigma v_{\rm rel} \rangle}{F(x_t/x) \,  x^2}.  
\label{Y:Boltzmann3}
\eea 
Mathematically, the braneworld cosmological effect is equivalent 
    to modifying the DM creation/annihilation cross section in the standard Big Bang cosmology as
\bea 
 \langle \sigma v_{\rm rel} \rangle \to 
 \left( \frac{\langle \sigma v_{\rm rel} \rangle}{F(x_t/x)} \right)  
 = \langle \sigma v_{\rm rel} \rangle 
   \left( \frac{x_t}{x}\right)^{-\gamma} 
\label{scale}   
\eea   
   for $ x_t/x > 1$.  
This equation implies that the braneworld cosmological effect 
 enhances (reduces) the DM relic abundance for $\gamma < 0$ ($\gamma > 0$).

In our RS and GB cosmologies, we can easily solve Eq.~(\ref{Y:Boltzmann3}) 
   with the cross section of  Eq.~(\ref{Xsec}). 
We then obtain  
\bea
  \Omega_{\rm DM} h^2  \simeq \frac{m \, Y(x=1) \, s_0}{\rho_c/h^2} 
  \simeq 1.16 \times 10^{24}  \, \frac{g_{DM}^2}{g_*^{3/2}}  \, g_V^4 \times \frac{x_t^{-\gamma}}{\gamma+1}. 
\label{Omega_BW}
\eea
The resultant DM density is modified by a factor of $R(\gamma)=\frac{x_t^{-\gamma}}{\gamma+1}$ 
   from the one in the standard Big Bang cosmology. 
Therefore, in order to reproduce the observed DM density of $\Omega_{\rm DM} h^2=0.12$, 
   the coupling $g_V$ is fixed to be 
\bea    
g_V = 2.31 \times 10^{-6}  \; R(\gamma)^{-1/4}.
\eea
For the RS and the GB cosmologies, we find 
   $R(2) \simeq 1.32 \, \sqrt{x_t}$ and $R(-2/3) \simeq 0.76 \, x_t^{-1/6}$, respectively. 
When $x_t \gg 1$, or equivalently, $m \gg T_t$, the braneworld cosmological effect 
   requires a different coupling value to reproduce the observed DM density.

\section{Application to $Z^\prime$-portal RHN DM with a light $Z^\prime$ boson}
\label{sec:4}
\begin{table}[t]
\begin{center}
  \begin{tabular}{|c|ccc|c|c|}
\hline  
    & $SU(3)_C$ & $SU(2)_L$ & $U(1)_Y$ & $U(1)_{B-L}$  & Z$_2$ \\ \hline
    $q_L^i$ & \bf{3} & \bf{2} & $1/6$ & $1/3$ & $+$ \\
    $u_R^i$ & \bf{3} & \bf{1} & $2/3$ & $1/3$ & $+$ \\
    $d_R^i$ & \bf{3} & \bf{1} & $-1/3$ & $1/3$ & $+$ \\ \hline
    $l_L^i$ & \bf{1} & \bf{2} & $-1/2$ & $-1$ & $+$ \\
    $N_R^j$ & \bf{1} & \bf{1} & $0$ & $-1$ & $+$ \\
    $N_R$ & \bf{1} & \bf{1} & $0$ & $-1$ & $-$ \\
    $e_R^i$ & \bf{1} & \bf{1} & $-1$ & $-1$ & $+$ \\ \hline
    $H$ & \bf{1} & \bf{2} & $-1/2$ & $0$ & $+$ \\
    $\Phi$ & \bf{1} & \bf{1} & $0$ & $2$ & $+$ \\
\hline    
  \end{tabular}
  \renewcommand{\baselinestretch}{1.1}
  \caption{
 The particle content of the minimal $B-L$ model with the RHN DM \cite{Okada:2010wd}. 
In addition to the SM particle content ($i = 1, 2, 3$ for three generations), 
   the three right-handed neutrinos categorized into 2+1 ($N_R^j$ ($j = 1, 2$) and $N_R$)
   and the $B-L$ Higgs field ($\Phi$) are introduced.
The $Z_2$-odd $N_R$ is a unique DM candidate in the model. 
} 
\label{Tab:1}  
  \end{center}
\end{table}

In the previous section, we have evaluated the freeze-in DM density with a light vector-boson mediator. 
We have found that the resultant relic density is independent 
  of the DM mass, and the observed relic density is reproduced by adjusting
  the coupling ($g_V$) of the light mediator.  
As we have found, the resultant DM density in the RS and GB braneworld cosmologies
  can be significantly altered from the one in the standard Big Bang cosmology. 
In other words, the $g_V$ value to reproduce $\Omega_{\rm DM} h^2=0.12$
   is changing in accordance with a braneworld model and the transition temperature $T_t$ (or equivalently, $x_t$). 
As we will discuss in the following, this coupling change has an impact on the search
   for the vector-boson mediator at future experiments. 
To see this, we consider a simple ``$Z^\prime$-portal DM'' scenario in this section.

The minimal $B-L$ model \cite{Davidson:1978pm, Mohapatra:1980qe, Marshak:1979fm,
          Wetterich:1981bx, Masiero:1982fi,  Mohapatra:1982xz, Buchmuller:1991ce} 
    is a simple, well-motivated model for the neutrino mass generation. 
In the model, the accidental U(1)$_{B-L}$ global symmetry of the SM is gauged,  
    and three RHNs are introduced to keep the model free from all gauge and mixed
    gauge-gravitational anomalies.     
The U(1)$_{B-L}$ symmetry is broken by the vacuum expectation value of a $B-L$ Higgs,
    which generates the $B-L$ gauge boson ($Z^\prime$) mass as well as Majorana masses 
    for the three RHNs. 
After the electroweak symmetry breaking, tiny neutrino masses are generated 
    through the type-I seesaw mechanism
    \cite{Minkowski:1977sc, Yanagida:1979as, GellMann:1980vs, Mohapatra:1979ia}.  
A concise way to incorporate a DM candidate into the minimal $B-L$ model 
   has been proposed in Ref.~\cite{Okada:2010wd}, 
   where instead of introducing a new particle, a $Z_2$ symmetry is introduced
   while keeping the minimal $B-L$ model particle content intact. 
We assign an odd-parity for one RHN, while all the other particles in the model are even. 
In this way, the parity-odd RHN is stable and serves as the DM in our universe. 
It is known that only two RHNs are necessary for a realistic seesaw scenario 
   to reproduce the neutrino oscillation data. 
This setup is called ``Minimal Seesaw" \cite{King:1999mb, Frampton:2002qc}. 
Therefore, through the $Z_2$-parity, the three RHNs are categorized into $2+1$ 
  with two RHNs for the seesaw mechanism and one RHN for the DM candidate. 
The particle content of the model is listed in Table \ref{Tab:1}. 
Except for the $Z_2$-parity assignment, the particle content is exactly the same as the one of the minimal $B-L$ model.

The RHN DM can communicate with the SM particles through the Higgs boson exchange (Higgs-portal) 
    and/or the $Z^\prime$ boson exchange ($Z^\prime$-portal). 
For a RHN DM as a thermal DM particle, 
  the Higgs-portal case \cite{Okada:2010wd, Okada:2012sg, Basak:2013cga}
  and the $Z^\prime$ portal case \cite{Okada:2016gsh, Okada:2016tci, Oda:2017kwl, Okada:2017dqs}
   (see \cite{Okada:2018ktp} for a review) have been extensibly studied.  
In the study for the $Z^\prime$-portal case, it has been pointed out that 
  the DM physics and the search for a $Z^\prime$ boson resonance 
  at the Large Hadron Collider (LHC) are complementary to narrow down the allowed model-parameter space.     
For a similar study of a $Z^\prime$-portal Dirac fermion DM in the context of the minimal $B-L$ model, 
   see  Refs.~\cite{FileviezPerez:2018toq,  FileviezPerez:2019cyn}. 
The RHN DM as a freeze-in DM particle has also been studied \cite{Kaneta:2016vkq, Biswas:2016bfo} 
  (For a similar study of the Dirac fermion case, see  Refs.~\cite{Heeba:2019jho, Mohapatra:2019ysk}). 
In particular, the $Z^\prime$-portal RHN DM with a light $Z^\prime$ boson has been studied 
   in Ref.~\cite{Kaneta:2016vkq} and it has been pointed out that the parameter region 
   motivated from the DM physics can be explored 
   by the planned/proposed experiments at the Lifetime Frontier.  
In the following, we extend the analysis in Ref.~\cite{Kaneta:2016vkq} 
   to the RS and GB braneworld cosmologies and investigate an impact of the braneworld cosmological
   effect on the Lifetime Frontier experiments.

In order to evaluate the freeze-in RHN DM relic density, 
   we first evaluate a thermal-averaged cross section for the RHN pair creation 
   from thermal plasma. 
The main process is  $f \bar{f} \to Z^\prime \to NN$ \cite{FileviezPerez:2018toq} and 
  its explicit form is given by
\bea
 \sigma(s)
 = \frac{13}{48 \pi} g_{BL}^4 
 \frac{\sqrt{s( s - 4m^2 )}}{s^2}, 
 \label{annihilation_X}
\eea
where $g_{BL}$ is the $B-L$ gauge coupling, and 
  we have neglected $Z^\prime$ boson mass ($m_{Z^\prime} \ll m$). 
The thermal average of the pair creation/annihilation cross section is calculated as 
\begin{eqnarray} 
 \langle \sigma v_{\rm rel} \rangle
 =  (s Y_{EQ})^{-2} \, g_{DM}^2 \frac{m}{64\pi^4 x} 
   \,  \int^{\infty}_{4 m^2} \, ds  \, 2(s-4m^2) \, \sigma(s) \, \sqrt{s} \, K_1 \left( \frac{x \sqrt{s}}{m} \right),
\end{eqnarray}
 where 
\begin{eqnarray}
 s Y_{EQ} =  \frac{g_{DM}}{2\pi^2} \frac{m^3}{x} K_2(x),
\end{eqnarray}
  $g_{DM}=2$, and $K_i$ is the modified Bessel function of the $i$-th kind.
For fixed values of $g_{BL}$ and $m$,  we obtain $\langle \sigma v_{\rm rel} \rangle$ as a function of $x$, 
  and numerically solve the Boltzmann equation to evaluate the relic density. 
As we have found in the previous section, the braneworld cosmological effects 
  are encoded in Eq.~(\ref{scale}) and we simply scale the thermal-averaged cross section by $(x/x_t)^\gamma$.

\begin{figure}[t]
\begin{center}
{\includegraphics*[width=0.7\linewidth]{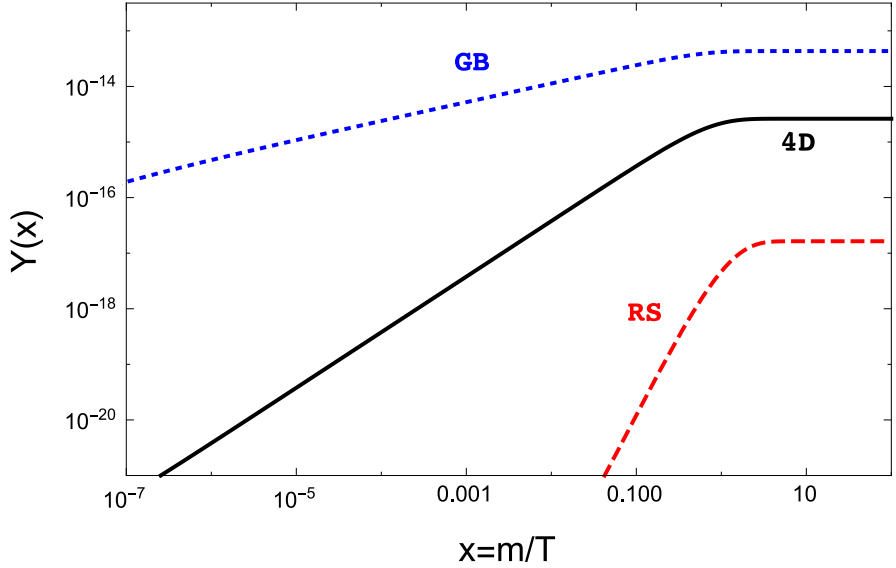}}
\caption{
Numerical solutions of the Boltzmann equation 
  in the GB braneworld cosmology (dotted blue line), 
  the Standard Big Bang cosmology (solid black line), 
  and the RS braneworld cosmology (dashed red line). 
Here, we set $m=10$ TeV, $T_t=1$ TeV, $m_{Z^\prime}=1$ GeV, and $g_{BL}=1.54 \times 10^{-6}$.  
}
\label{Fig:1}
\end{center}
\end{figure}

For example values of the input parameters ($m=10$ TeV, $T_t=1$ TeV, $m_{Z^\prime}=1$ GeV, 
   and $g_{BL}=1.54 \times 10^{-6}$), 
   we show in Fig.~\ref{Fig:1} the numerical solutions of the Boltzmann equation 
   in the standard Big Bang (black solid line), RS (red dashed line) and GB (blue dotted line) cosmologies. 
We find that as long as $m \gg m_{Z^\prime}$, the results are independent of $m$ and $m_{Z^\prime}$.   
As we see in the figure, the resultant DM density is enhanced (reduced) 
   in the GB (RS) cosmology, compared to the one obtained in the standard cosmology.     
Corresponding DM relic densities are  
  $\Omega_{\rm DM} h^2=7.2 \times 10^{-3}$ for the standard cosmology 
  while $\Omega_{\rm DM} h^2=0.12$ and $4.5 \times 10^{-5}$ for the GB and RS cosmologies,
  respectively.  
We can see  that Eq.~(\ref{Omega_BW}) is satisfactory as a rough estimate. 
The discrepancy between the formula of Eq.~(\ref{Omega_BW}) and the actual numerical solution
   originates from the fact that the thermal-averaged cross section is not exactly proportional to $x^2$ 
   around $x \sim 1$ while it is a very good approximation for $x \ll 1$. 

\begin{figure}[t]
\begin{center}
{\includegraphics*[width=0.6\linewidth]{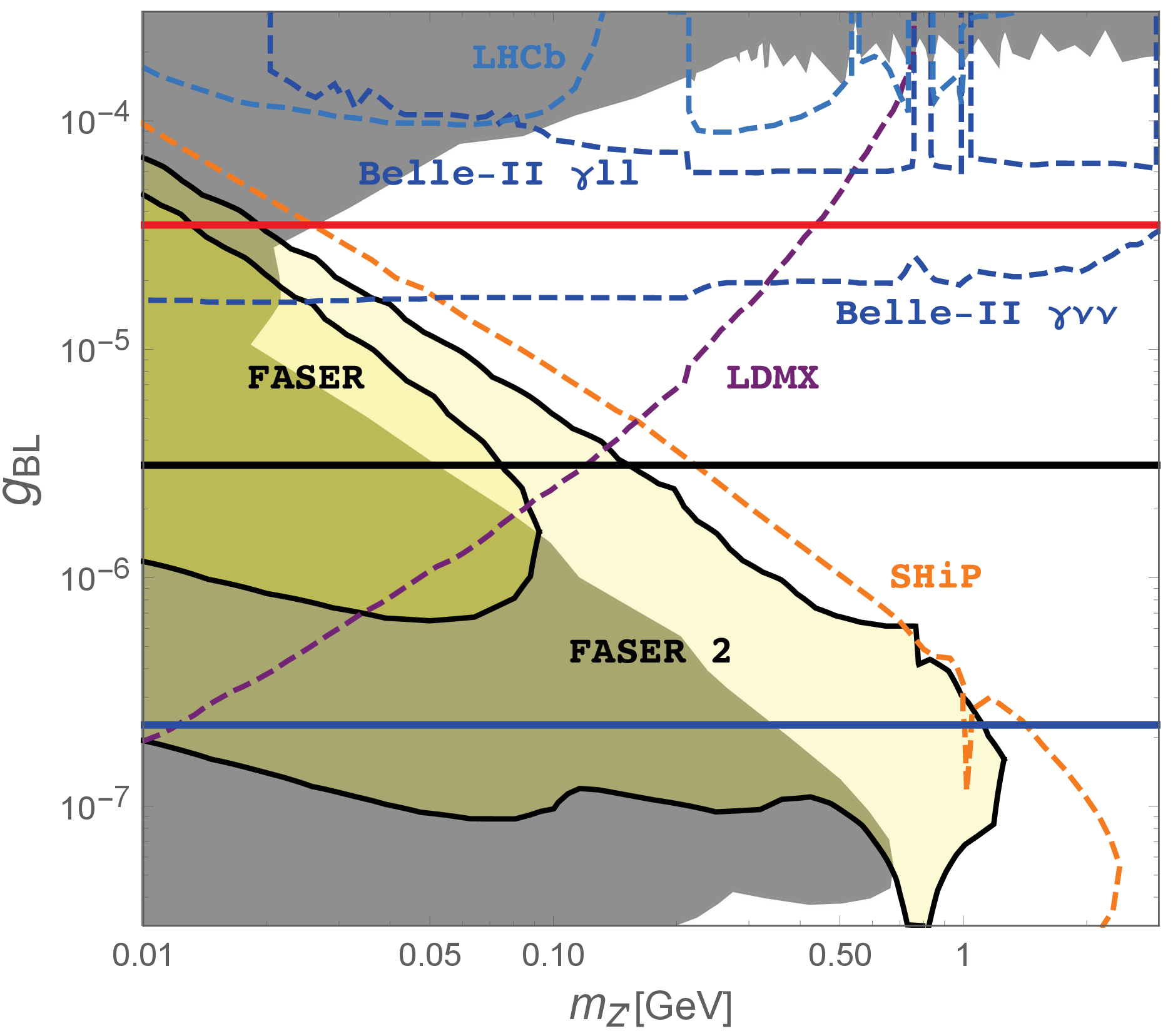}}
\caption{
Gauge coupling values as a function of $m_{Z^\prime} \ll m$ to reproduce the observed DM relic density 
   in the standard Big Bang (horizontal black line), RS (horizontal red line) and GB (horizontal blue line) cosmologies,  
   along with the search reach of various planned/proposed experiments at the Lifetime Frontier 
   and the current excluded region (gray shaded). 
We have fixed $x_t=100$  ($x_t=10^6$) for the RS (GB) cosmology.  
the gauge couplings are found to be 
  $g_{BL}=3.50 \times10^{-5}$, $3.11 \times 10^{-6}$ and $2.27 \times 10^{-7}$ 
  for the RS, standard Big Bang and GB cosmologies, respectively.  
}
\label{Fig:2}
\end{center}
\end{figure}

Let us now discuss an impact of the braneworld cosmological effect on the future experiments at the Lifetime Frontier. 
In order to reproduce the observed relic density for the RHN DM via the light $Z^\prime$-portal interaction, 
   the $B-L$ gauge coupling is found to be very small. 
Hence, the $Z^\prime$ boson becomes long-lived. 
Such a long-lived $B-L$ gauge boson can be explored at Lifetime Frontier experiments. 
The recently approved ForwArd Search Experiment (FASER) \cite{Feng:2017vli, Ariga:2018uku, Ariga:2019ufm} 
    has a physics run planned at the LHC Run-3 and  its upgraded version (FASER 2) at the High-Luminosity LHC. 
The prospect of the $B-L$ gauge boson search at FASER is summarized in Ref.~\cite{Ariga:2018uku}.  
FASER 2 can search for a long-lived $Z^\prime$ boson 
   with its mass in the range of 10 MeV$\lesssim m_{Z^\prime} \lesssim 1$ GeV
   for the $B-L$ gauge coupling in the range of $10^{-8}  \lesssim g_{BL} \lesssim 10^{-4.5}$. 
The planned/proposed experiments, 
   such as Belle II \cite{Dolan:2017osp}, LHCb \cite{Ilten:2015hya, Ilten:2016tkc}, SHiP \cite{Alekhin:2015byh}
   and LDMX \cite{Berlin:2018bsc}, 
   which will also search for a long-lived $Z^\prime$ boson,  
   will cover a parameter region complementary to FASER.

In Fig.~\ref{Fig:2}, we show our results for the $B-L$ gauge coupling as a function of $m_{Z^\prime}$ 
   to reproduce the observed DM relic density in the standard Big Bang (horizontal black line), 
   RS (horizontal red line) and GB (horizontal blue line) cosmologies,  
   along with the search reach of various planned/proposed experiments at the Lifetime Frontier 
   and the current excluded region (gray shaded) \cite{Bauer:2018onh}. 
Here, we have fixed $x_t=100$  ($x_t=10^6$) for the RS (GB) cosmology.  
As expected, we find that the results are independent of $m_{Z^\prime} \ll m$.    
The braneworld effects shift the resultant $g_{BL}$ value upwards in the RS cosmology 
  and downwards in the GB cosmology from the one in the standard cosmology. 
Therefore, if a long-lived $Z^\prime$ boson is observed in the future, 
  we can measure not only $g_{BL}$ and $m_{Z^\prime}$ but also obtain 
  the information about possible non-standard evolution of the early universe.

\section{Conclusions}
\label{sec:5}
In the 5D braneworld cosmology, the Friedmann equation of our 4D universe on a brane 
  is modified at high temperatures while the standard Big Bang cosmology 
  is recovered at low temperatures. 
Based on two well-known scenarios,  
  the Randall-Sundrum and Gauss-Bonnet braneworld cosmologies, 
  we have investigated the braneworld cosmological effect on the relic density 
  of a non-thermal DM particle 
  whose interactions with the Standard Model particles are so weak that 
  its relic density is determined by the freeze-in mechanism. 
For DM production processes in the early universe,    
  we have considered a simple DM scenario with a light mediator for the DM particle
  to communicate with the Standard Model particles. 
We have found that the braneworld cosmological effect can dramatically alter 
  the resultant DM density from the one in the standard Big Bang cosmology.  
As an application, we consider the $Z^\prime$-portal RHN DM in the minimal $B-L$ extended Standard Model 
  with a light $Z^\prime$ boson as a mediator. 
We have found an impact of the braneworld cosmological effect 
  on the search for the long-lived $Z^\prime$ boson at the planned/proposed Lifetime Frontier experiments.

\section*{Acknowledgments}
The author V.B. would like to thank the University of Alabama and the McNair Scholars Program 
   for support in the form of the McNair Fellowship. 
This work is supported in part by 
  the United States Department of Energy Grant DE-SC-0012447 (N.O.),  
  and the M. Hildred Blewett Fellowship of the American Physical Society, www.aps.org (S.O.).

\bibliographystyle{utphysII}
\bibliography{References}

\end{document}